PROCEEDINGS OF THE XIII FEOFILOV SYMPOSIUM
"SPECTROSCOPY OF CRYSTALS DOPED
BY RARE-EARTH AND TRANSITION-METAL IONS"
(Irkutsk, July 9–13, 2007)

# Location of the Energy Levels of the Rare-Earth Ion in $BaF_2$ and $CdF_2$

P. A. Rodnyĭ[a], I. V. Khodyuk[a], and G. B. Stryganyuk[a]

[a] *St. Petersburg State Polytechnical University, Politekhnicheskaya ul. 29, St. Petersburg, 195251 Russia*
e-mail: Khodyuk@tuexph.stu.neva.ru
[b] *Franko Lviv National University, ul. Dragomanova 50, Lviv, 79005 Ukraine*

**Abstract**—The location of the energy levels of rare-earth (*RE*) elements in the energy band diagram of $BaF_2$ and $CdF_2$ crystals is determined. The role of $RE^{3+}$ and $RE^{2+}$ ions in the capture of charge carriers, luminescence, and the formation of radiation defects is evaluated. It is shown that the substantial difference in the luminescence properties of $BaF_2 : RE$ and $CdF_2 : RE$ is associated with the location of the excited energy levels in the band diagram of the crystals.

PACS numbers: 61.72.Ww, 76.30.Kg, 73.20.Hb

**DOI:** 10.1134/S1063783408090072

## 1. INTRODUCTION

The location of energy levels of rare-earth (*RE*) elements in various matrices plays an important role in the physical processes occurring in crystals. In recent years, owing to the generalization of a great amount of data available in the literature on different characteristics of rare-earth elements in ionic compounds [1–3], it has become possible to determine the location of the ground and excited levels of these rare-earth elements in the energy level diagram of some crystals. In this study, we determined the location of energy levels of rare-earth elements in the band diagram for crystals of two scintillators $BaF_2$ and $CdF_2$, which exhibit intrinsic luminescence, and evaluated the effect of rare-earth elements on some physical properties of these objects, specifically on the charge carrier capture.

The location of the ground and excited energy levels of rare-earth states exerts a noticeable effect on the luminescence properties of crystals. For example, the $CdF_2$ crystal exhibit intrinsic luminescence, which is substantially quenched at room temperature [4]. Repeated attempts to increase the intensity of luminescence of the $CdF_2$ crystal by means of the introduction of an activator, including $Ce^{3+}$, have failed. In this study, we attempted to answer the question as to why this rare-earth element serves as a good activator, i.e., produces intense luminescence, in one matrix and does not luminesce at all in another matrix.

Rare-earth elements embedded in halides and oxides can affect their radiation resistance, which is particularly important for scintillation crystals. In some cases, a small amount of impurities substantially increases the radiation resistance (the permissible dose) of the crystal. For example, the permissible dose for the $Gd_2SiO_5$ : Ce (0.5 at %) scintillator is several orders of magnitude higher than that for pure $Gd_2SiO_5$ [5]. The physical mechanism of enhancement of the radiation resistance is poorly understood, and the existing models of the process have a number of contradictions. It was assumed that the $Ce^{3+}$ and $Pr^{3+}$ ions, which exhibit a tendency toward transformation into the tetravalent state under irradiation, exert a negative effect on the radiation resistance of crystals. At the same time, the Eu, Sm, and Yb ions, which change the charge from +3 to +2, should suppress the formation of radiation defects [6]. The above change in the charge state of the rare-earth elements should undoubtedly be taken into account; however, this change cannot be considered as a basic phenomenon, if for no other reason than the fact that an increase in the radiation resistance upon introduction of $Ce^{3+}$ ions has been observed for a number of crystals and glasses [5, 7]. Another idea is that the introduction of trivalent rare-earth elements into crystals with divalent cations of the host matrix ($BaF_2$, $CdF_2$) decreases the amount of anion vacancies in the crystal. This circumstance, in turn, decreases the concentration of *F* centers [8]. However, it is known that the initial vacancies bring about the effective formation of *F* centers only at the early stage of the process. In this paper, we propose the use of a new approach to solving the problem of radiation resistance of crystals, which is based on the location of the energy levels of $RE^{3+}$ ions in $BaF_2$ and $CdF_2$.

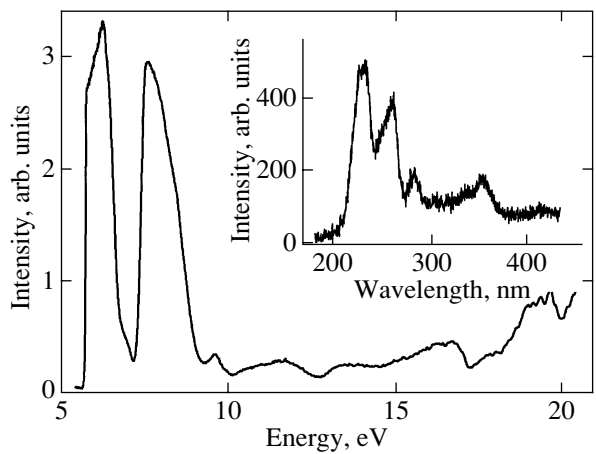

**Fig. 1.** Excitation spectrum of the luminescence band at 260 nm for the $BaF_2 : Pr^{3+}$ (0.3 at %) crystal at room temperature. The inset shows the luminescence spectrum of the crystal upon excitation by photons with an energy of 6.1 eV.

## 2. SAMPLE PREPARATION AND EXPERIMENTAL TECHNIQUE

The excitation and emission spectra of crystals were measured at room temperature at the HASYLAB Synchrotron Research Laboratory (DESY, Hamburg) with the use of experimental equipment of the SUPERLUMI station. The measurement of the luminescence spectra was performed under continuous x-ray excitation, and the kinetics was measured under pulsed x-ray excitation. The experimental technique was described in more details in [4, 9]. Crystals of the $BaF_2 : RE$ and $CdF_2 : RE$ were grown by the Stepanov–Stockbarger method at the Vavilov State Optical Institute (St. Petersburg, Russia).

## 3. RESULTS AND DISCUSSION

Figure 1 shows the excitation spectrum of the UV (250-nm) luminescence band of the $BaF_2 : Pr^{3+}$ crystal. This luminescence band can undoubtedly be attributed to the $d$–$f$ interconfiguration transitions of the $Pr^{3+}$ ion both in the spectral position (inset to Fig. 1) and in the decay time (22 ns). The excitation spectrum involves two characteristic bands with maxima at energies of 6.1 and 7.5 eV. The low-energy absorption edge (5.5 eV) allows one to determine the location of the $5d$ level with respect to the $4f$ electron configuration in the ground state of the $Pr^{3+}$ ion. The corresponding absorption edge for $Ce^{3+}$ and $BaF_2 : Ce^{3+}$ is located at an energy of 4.1 eV.

Figures 2 and 3 present the energy level diagrams for the $BaF_2 : RE$ and $CdF_2 : RE$ crystals, respectively. When constructing these diagrams, we used the following data: the luminescence excitation spectra of the $Pr^{3+}$ and $Ce^{3+}$ ions in the $BaF_2$ crystal, the previously measured excitation spectra of rare-earth elements in $BaF_2$ [9–11], the data available in the literature on the photoconductivity and photoionization of rare-earth elements in crystals [12–14], and a number of model–theoretical considerations [1, 2]. The results obtained from the generalization of a large amount of data available in the literature for rare-earth elements in crystals [1–3] can be summarized as follows.

(1) The relative energy position of the ground ($4f$) states of the $RE^{3+}$ and $RE^{2+}$ ions remains almost unchanged in the series of lanthanides and weakly depends on the nature of the matrix into which these ions were embedded.

(2) The energy position of the excited $5d$ levels of the $RE^{3+}$ and $RE^{2+}$ ions is determined by the environment (the crystal field) and weakly depends on the type of the ion itself.

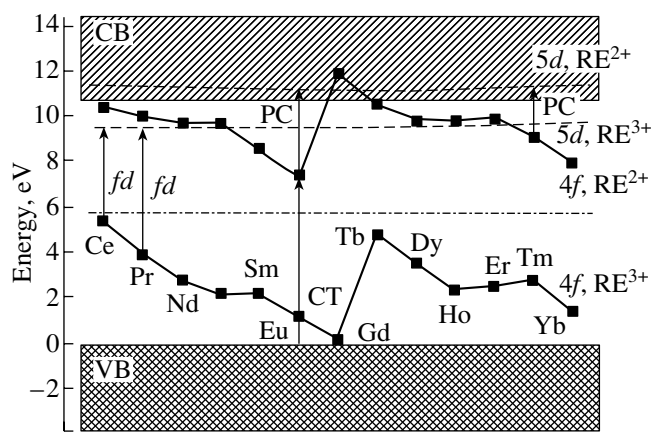

**Fig. 2.** Energy level diagram of the rare-earth ions in the $BaF_2$ crystal ($RE^{3+}$ $4f$ and $RE^{2+}$ $4f$ are the ground states, and $RE^{3+}$ $5d$ and $RE^{2+}$ $5d$ are the lower excited states). The dash-dotted line indicates the Fermi level $E_F$. Designations of the electron transitions: PC is photoconductivity, CT is the charge transfer, and $fd$ is the low-energy edge of the absorption band for the $4f$–$5d$ transitions.

(3) The energy gap between the $RE^{3+}$ 5d and $RE^{2+}$ 5d levels for the particular matrix (crystal) remains nearly constant.

(4) In charge-transfer transitions of the $RE^{3+}$ ions, the initial state is associated with the top of the valence band, whereas the final state coincides with the ground state of the corresponding $RE^{3+}$ ion. As a result, the maximum of the absorption band upon charge transfer of the $RE^{3+}$ ion unambiguously determines the position of the 4f level of the corresponding $RE^{2+}$ ion with respect to the valence band.

Therefore, it is sufficient to know the energy parameters of one rare-earth element in order to construct the energy level diagram for all rare-earth elements. Since the particularly large amount of data are available on the energy position of the charge-transfer band of $Eu^{3+}$ ions in different matrices [3], europium usually serves as a reference point in the construction of energy level diagrams similar to those presented in Figs. 2 and 3. When constructing the scheme depicted in Fig. 2, we used the data on the photoconductivity of the $BaF_2$ : $Eu^{2+}$ [12], $BaF_2$ : $Tm^{2+}$ [13], and $BaF_2$ : $Ce^{3+}$ [14] crystals. According to the data reported in [12], the lower 5d level of the $Eu^{2+}$ ion is located ~0.4 eV above the bottom of the conduction band. For the $BaF_2$ : $Ce^{3+}$ crystal, we took into account the approximate value of the energy of the charge-transfer band, which was indirectly estimated in [15]. The location of the fundamental 4f level of the $Ce^{2+}$ ion should be discussed separately. As follows from the theory, this level should be located in the conduction band of the $BaF_2$ crystal [1]. However, according to [14], the photoconductivity threshold for $Ce^{2+}$ ions in $BaF_2$ is equal to 1.1 eV. The point is that the ground state of the $Ce^{2+}$ ion is $4f5d$ rather than $4f^2$; in addition, $Ce^{2+}$ is the largest ion among the $RE^{2+}$ ions, which increases the lattice distortion near the ion.

When constructing the energy level diagram of the $CdF_2$ : $RE^{2+}$ crystal, the energy position of the $Eu^{2+}$ 4f level (0.35 eV) with respect to the bottom of the conduction band [16] and a number of spectral characteristics [17, 18] served as the main reference point.

The energy levels of the rare-earth elements determine the state of the electron and hole traps in the crystal. Trivalent ions with the ground (4f) states located above the top of the valence band can capture holes from the valence band. Correspondingly, the $RE^{2+}$ ions with the 4f levels located below the bottom of the conduction band can capture electrons from the conduction band. The energy levels of the rare-earth ions, which are located near the bottom of the conduction band ($\Delta E < 1$ eV), serve as shallow-level electron traps. These centers are thermally unstable and responsible for the undesirable persistent luminescence of scintillators. The impurities with $\Delta E > 1$ eV serve as deep (stable)

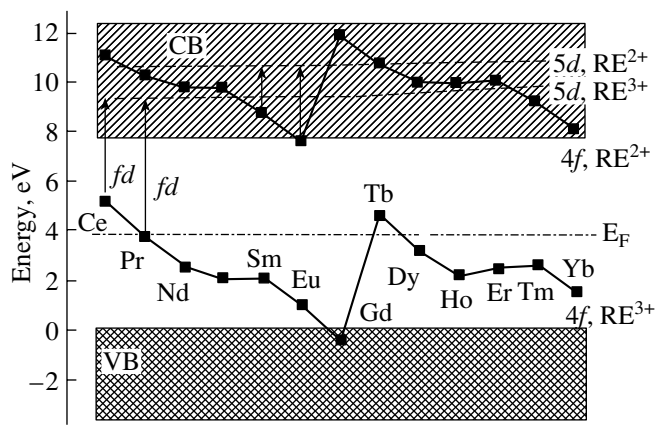

**Fig. 3.** The same as in Fig. 2 for the $CdF_2$ crystal.

carrier traps; moreover, they are important for phosphors with optical memory [19].

In barium fluoride, the $Eu^{3+}$, $Sm^{3+}$, and $Yb^{3+}$ ions are deep electron traps, because, after the electron capture, they are transformed into $Eu^{2+}$, $Sm^{2+}$, and $Yb^{2+}$ ions with the ground levels located 1–3 eV below the conduction band (Fig. 2). The $Nd^{3+}$, $Ho^{3+}$, and $Er^{3+}$ ions are shallow electron traps, whereas the $Pr^{3+}$, $Nd^{3+}$, $Tb^{3+}$, and $Dy^{3+}$ ions should be considered to be deep hole traps.

For the $CdF_2$ : $RE$ crystal, the situation differs from that for the $BaF_2$ : $RE$ crystal. Trivalent ions (Eu, Sm, and Yb), which exhibit a tendency toward a transition to the divalent state, cannot be electron traps in the $CdF_2$ crystal because the corresponding $RE^{2+}$ 4f states are located in the conduction band (Fig. 3). In this case, hole traps dominate; in particular, the $Nd^{3+}$ and $Dy^{3+}$ ions should be considered to be deep hole traps. It is known that the $CdF_2$ compound containing rare-earth ions (except for europium) can be transformed into the semiconductor state by means of its annealing in cadmium vapors [20]. We assume that this property of the $CdF_2$ : $RE^{2+}$ compound is determined by the energy location of the $RE^{2+}$ 4f levels (except for the $Eu^{2+}$ 4f level) in the conduction band of the crystal (Fig. 3).

The energy diagrams constructed in this study are also very useful in evaluating the luminescence properties of rare-earth elements in the crystal under consideration. It can be seen from Fig. 2 that all the excited 5d states of divalent ions lie in the conduction band of the $BaF_2$ crystal. The states localized inside the conduction band effectively interact (intermix) with band states; consequently, the d–f luminescence of $RE^{2+}$ ions is impossible in $BaF_2$. This inference was confirmed in the experiment: the d–f luminescence in $BaF_2$ : $RE^{2+}$ was not observed ($BaF_2$ : $Eu^{2+}$ exhibits a so-called anomalous luminescence, whereas $BaF_2$ : $Sm^{2+}$ is characterized by transitions from the $Sm^{2+}$ $^5D_0$, 4f level

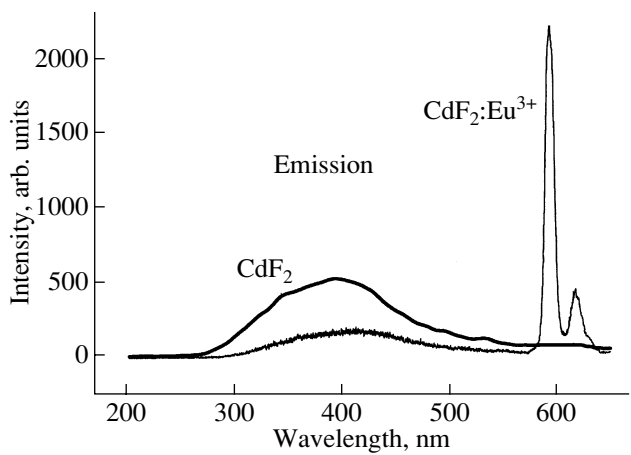

**Fig. 4.** Luminescence spectra of the pure $CdF_2$ crystal and the $CdF_2$ : $Eu^{3+}$ crystal upon x-ray excitation. $T$ = 300 K.

which is located below the 5d levels and slightly below the bottom of the conduction band).

In the $CdF_2$ crystal, all the excited levels of the mixed $4f5d$ configuration of the $RE^{3+}$ and $RE^{2+}$ ions are located in the conduction band. The exclusion is the $CdF_2$ : $Eu^{3+}$ crystal, which luminesces (Fig. 4) as resulting from the fact that the charge-transfer band is partially overlapped with the conduction band. As can be seen from Fig. 4, the f–f luminescence of the $Eu^{3+}$ ion (the $^5D_0 \longrightarrow {}^7F_j$ transitions) suppresses intrinsic emission (the broad band at 400 nm) of the $CdF_2$ crystal.

The shift of the 5d levels to the conduction band in the series of $CaF_2$ [2], $BaF_2$, and $CdF_2$ is associated primarily with an increase in the covalency of the crystals. It seems likely that the absence of luminescence of the rare-earth elements in the $PbF_2$ crystal (with a band gap of 5.84 eV) is also associated with the arrangement of the excited 5d levels of the $RE^{3+}$ and $RE^{2+}$ ions in the conduction band.

The location of the energy levels of the rare-earth elements should be taken into account when analyzing the mechanism of defect formation in crystals. It is known that halides are characterized by the effective axial relaxation of anions, which leads to the formation of $V_k$ centers. Since the crystal structure is distorted in the vicinity of the $V_k$ center, this state has a tendency toward the formation of point defects. It is clear that the retardation of defect formation in the crystal can be achieved by means of the introduction of impurities (including rare-earth elements) involved in the process of capture of valence holes, which competes with the process of the formation of $V_k$ centers.

According to the energy level diagram depicted in Fig. 3, virtually any $RE^{3+}$ ion (except for $Cd^{3+}$) in the $CdF_2$ crystal can be a hole trap and can prevent the formation of $V_k$ centers in the crystal. This inference corresponds to the experimental data: upon introduction of a small amount (0.5 at %) of $RE^{3+}$ ions into the $CdF_2$ crystal, the permissible dose of irradiation of the crystal increased, in particular, by a factor of $10^3$ for the $Nd^{3+}$ and $Sm^{3+}$ ions and by a factor of $10^5$ for the $Ce^{3+}$ and $Tb^{3+}$ ions [21].

In the $BaF_2$ crystal, the deep donors are $Ce^{3+}$, $Pr^{3+}$, and $Tb^{3+}$ ions, which could improve the radiation resistance of the crystals. However, the experiments showed that $Ce^{3+}$ ions deteriorate the radiative properties of the $BaF_2$ crystal [21]. The $Ce^{4+}$ ions effectively capture conduction electrons and return to the trivalent state after irradiation. A somewhat increase in the radiation resistance is observed upon introduction of ytterbium [21], which can serve as a trap for both holes and electrons.

## 4. CONCLUSIONS

Thus, the energy level diagram of rare-earth elements in $BaF_2$ and $CaF_2$ was constructed using the obtained results and data available in the literature. The role of $RE^{3+}$ and $RE^{2+}$ ions in the capture of charge carriers, luminescence, and the generation of radiation defects was determined. The substantial difference in the luminescence properties of the $BaF_2$ : $RE$ and $CdF_2$ : $RE$ crystals is associated with the location of the energy levels in the conduction band of the crystals. The radiation resistance of the crystals depends on the location of the $RE^{3+}$ 4f levels in the band gap of these materials.

Energy level diagrams similar to those presented in Figs. 2 and 3 can be constructed for any crystal with a sufficient number of known parameters. These diagrams are useful both for understanding the mechanisms of capture of electrons and holes and for investigating the emission and absorption spectra of crystals. To date, the energy level diagrams of rare-earth elements have been constructed for a number of oxides and sulfides [22].

*Translated by N. Korovin*